\documentclass[aps,twocolumn,showpacs,amsmath,amssymb,pre,superscriptaddress, floatfix]{revtex4-1}

\usepackage{graphicx}% Include figure files
\usepackage{dcolumn}% Align table columns on decimal point
\usepackage{bm}% bold math
\usepackage{amssymb}
\usepackage{multirow}

\usepackage{subfigure}

\begin{document}

\title{Analysis of interface conversion processes of ballistic and diffusive motion in driven superlattices}

\date{\today}


\author{Thomas Wulf}
\email[]{Thomas.Wulf@physnet.uni-hamburg.de}
\affiliation{Zentrum f\"ur Optische Quantentechnologien, Universit\"at Hamburg, Luruper Chaussee 149, 22761 Hamburg, Germany}%
\author{Christoph Petri}
\affiliation{Zentrum f\"ur Optische Quantentechnologien, Universit\"at Hamburg, Luruper Chaussee 149, 22761 Hamburg, Germany}%
\author{Benno Liebchen}
\affiliation{Zentrum f\"ur Optische Quantentechnologien, Universit\"at Hamburg, Luruper Chaussee 149, 22761 Hamburg, Germany}%
\author{Peter Schmelcher}
\email[]{Peter.Schmelcher@physnet.uni-hamburg.de}
\affiliation{Zentrum f\"ur Optische Quantentechnologien, Universit\"at Hamburg, Luruper Chaussee 149, 22761 Hamburg, Germany}%

\begin{abstract}

We explore the non-equilibrium dynamics of non-interacting classical particles in a one-dimensional driven superlattice which is composed of domains exposed 
to different time-dependent forces.
It is shown how the combination of directed transport and conversion processes from diffusive to ballistic motion  
causes strong correlations between velocity and phase for particles passing through a superlattice. 
A detailed understanding of the underlying mechanism 
allows us to tune the resulting velocity distributions at distinguished points in the superlattice by means of local variations of the applied driving force.
As an intriguing application we present a scheme how initially diffusive particles can be transformed into a monoenergetic pulsed particle beam whose parameters such as its energy can be varied.

\end{abstract}

\maketitle
%%%%%%%%%%%%%%%%%%%%%%%%%%%%%%%%%%%%%%%

\section{Introduction}

Dynamical systems and their transport properties have been studied extensively over the last decades \cite{Wiggins:1992}, whereas the topic 
stimulated a vast amount of research when the possibility of directed currents in the absence of a mean force was realised \cite{Maddox:1993, Bartussek:1994, Dykman:1997}.
Since the second law of thermodynamics forbids
such transport phenomena in equilibrium, these systems have to be constantly driven out of equilibrium.
Early works \cite{Maddox:1993, Bartussek:1994, Dykman:1997} were based on
noise, i.e. statistical external fields, in combination with spatially asymmetric so called 'ratchet' potentials to overcome the limitations formulated by the second law of thermodynamics and thus to evoke a particle current.
These type of systems are of particular interest because they outline a working principal for 
biological systems such as molecular motors \cite{Magnasco:1993, Juelicher:1997} or quantum motors \cite{Ponomarev:2009}.\\ 
However, it was soon realised that directed
currents can very well be obtained with deterministic external fields, as long as certain spatial- and temporal symmetries in the equations of motion are broken \cite{Flach:2000}, which was
investigated afterwards in a vast amount of literature (see \cite{Dittrich:2000, Denisov:2001, Yevtushenko:2001, Denisov:2002, Denisov:2006, Gong:2006, Wang:2007} and references therein).
These deterministic ratchets are of particular interest since they might have remarkable applications in nanoscale devices such as electron pumps or transistors \cite{Linke:2002}.
First experimental realisations 
included semiconductors or semiconductor micro structures where a combination of Laser fields has been applied
which led to directed currents in electron ratchets \cite{Alekseev:1999, Linke:1999}.
Directed currents also became a subject of interest in experiments concerning cold atoms in optical lattices \cite{Gommers:2005, Renzoni:2006, Salger:2009, Renzoni:2012}, where additional AC forces are applied
to drive the system out of equilibrium and at the same time break the required symmetries that would otherwise prevent transport phenomena. 
These type of experiments are  
of particular interest since they allow for a precise control over the system parameters and provide extensive tuneability in the used AC drivings \cite{Renzoni:2006, Renzoni:2012}. \\
While the so far mentioned works focus on only time-dependent AC forces,
it was found recently that a spatial dependence of these AC forces leads to a diversity of dynamical phenomena \cite{Petri:2010, Petri:2011, Liebche:2011}.
The latter studies address the classical dynamics of particles in a lattice with a site-dependent driving.
In \cite{Petri:2010} it is demonstrated how a phase-modulated lattice allows for directed transport even though the driving of each barrier on its own does not break the relevant symmetries \cite{Flach:2000}. 
Ref. \cite{Liebche:2011} shows how a ramping of the potential height in combination with a site-dependent driving leads to a patterned deposition of particles.
A specific realisation of a site-dependently driven lattice is the one of a block lattice (BL) as introduced in \cite{Petri:2011}, which is reminiscent of semiconductor heterostructures and superlattices. Indeed
only recently 
the possibility of ratchet effects in superlattices of semiconductor heterostructures with a superimposed periodic potential was reported \cite{Ganichev:2011, Ivchenko:2011}. 
In the case of \cite{Petri:2011}, the superlattice consists of different blocks containing many inidvidual barriers where the barriers of
each block are governed by a certain time-dependence i.e. driving law, whereas different blocks exhibit in general different driving laws. 
The long time transient dynamics in such a superlattice shows intriguing phenomena like the formation of spatial density oscillations. The latter
were explained and analysed by means of conversion processes from diffusive to ballistic (and vice versa) motion at the positions where two neighbouring blocks connect, i.e. at the interfaces of two blocks.  
However, a rigorous discussion of the processes occuring in a unit cell of such a superlattice -that is a system containing only two blocks each equipped with one of the used driving laws- is still missing and is 
therefore subject of the present work. 
In this sense we investigate the diffusive- to ballistic motion conversion processes in detail and explore their influence on the dynamics of particles leaving the two block system. 
As a result we obtain peaked velocity distributions for outgoing particles even though their initial conditions are chosen exclusively within the chaotic sea of the underlying phase space.
By adjusting parameters in the driving we are able to manipulate these velocity distributions in a controlled manner. Finally, we demonstrate how 
the insights gained from the two block system enable us to exploit the conversion processes in superlattices build up of many blocks each equipped with a unique driving law. 
In doing so an 
initially diffusive particle ensembles is converted into an ensemble with a velocity distribution containing a single peak, whereas both the width as well as the peaks mean velocity can be tuned. \\
The present work is structured in the following way: In section \ref{section2} we introduce the setup of a block lattice (BL). In section \ref{section3} the dynamics of a single block is discussed. 
We explore the conversion of diffusive to ballistic motion at distinguished positions in the BL in section \ref{section:4}. Additionally, the influence of these processes for outgoing particles in a simple 
two block system is discussed in section \ref{section:5}. Finally, we investigate 
the dynamics of superlattices containing several hundred blocks in section \ref{section:6}. Section \ref{section:7} contains our brief conclusions.

%%%%%%%%%%%%%%%%%%%%%%%%%%%%%%%%%%%%%%%%%%%%%%%%%%%%%%%%%%%%%%%%%%%%%%%%%%%%%%%%%%%%%%%%%%%%%%%%%%%%%%%%%%%%%%%%%%%%%%%%%%%%%%%%%%%%%%%%%%%%%%%%%%%%%%%%%%%%%%%%%%%%%%%%%%%%%%%%%%%%%%%%%%%%%%%%%
\section{Setup and Hamiltonian}
\label{section2}

\begin{figure}[htbp]
\centering
\includegraphics[width=0.9\columnwidth]{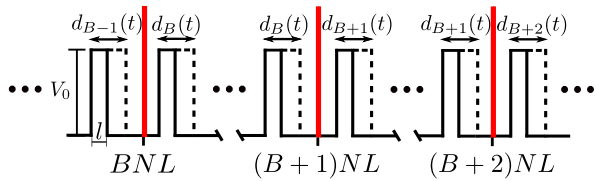}
\caption{\label{fig:lattice}Sketch of a lattice that consists of differently laterally driven blocks, where the barriers with $NB\leq i<(B+1)N$ are equipped with the same driving law $d_B(t)$.}
\end{figure}

The system investigated is a one-dimensional driven lattice consisting of laterally oscillating 
square potential barriers of equal height $V_0$ and length $l$ as sketched in Fig. \ref{fig:lattice}. 
Each barrier is characterised uniquely by its index $i$. 

Furthermore, 
the barrier positions are time-dependent and described by the so-called 'driving law' $d(t)$, which is chosen such that the $i$-th barrier remains at all times within an interval
of length $L$ expanding from $iL$ to $(i+1)L$. Additionally, the lattice is divided into blocks such that 
different driving laws $d_B(t)$ (introducing the 'block index' $B$) are used. In doing so, each $d_B(t)$  governs the barrier motions for the sites $NB\leq i<(B+1)N$ where $N$ denotes the number of barriers within one block
and is set to $10^4$ throughout this work. 
The general structure of the driving law is a biharmonic function $d_B(t)=A_B[\cos (\omega_B t) + \sin (2 \omega_B t + \Delta \Phi_B)]$ with three parameters 
$A_B$, $\omega_B$ and $\Delta \Phi_B$ which depend on the block index $B$.
Hence, the resulting classical Hamiltonian for noninteracting particles is given by 

\begin{equation}
 \begin{split}
  H(x,p,t)=\frac{p^2}{2m}+ \ \  \ \ \ \ \ \ \ \ \ \  \ \ \ \ \ \ \ \ \ \ \ \ \ \ \ \ \ \ \ \ \ \ \ \   \\
      \sum_{B=0}^{B=N_{\text{Bl}}-1}  \sum_{i=BN}^{(B+1)N} V_0 \Theta (l/2-|x-X_{0,i}-d_{B}(t)|),
  \end{split}
  \label{Hamiltonian}
\end{equation}

with $N_{\text{Bl}}$ being the number of considered blocks and $X_{0,i}$ the equilibrium position of the $i$-th barrier (chosen such 
that the barrier oscillates symmetrically within its unit cell, i.e. $\text{min}|X_{0,i}+d_B(t)-iL| = \text{min}|X_{0,i}+d_B(t)-(i+1)L|$).
Additionally, we set the mass $m=1$ without loss of generality and keep $V_0=1.0$, $L=5.0$ and $l=1.0$ constant throughout this work.\\

%%%%%%%%%%%%%%%%%%%%%%%%%%%%%%%%%%%%%%%%%%%%%%%%%%%%%%%%%%%%%%%%%%%%%%%%%%%%%%%%%%%%%%%%%%%%%%%%%%%%%%%%%%%%%%%%%%%%%%%%%%%%%%%%%%%%%%%%%%%%%%%%%%%%%%%%%%%%%%%%%%%%%%%%%%%%%%%%%%%%%%%%%%%%%%%%%
\section{Dynamics in a single block}
\label{section3}

\begin{figure}[htbp]
\centering
\includegraphics{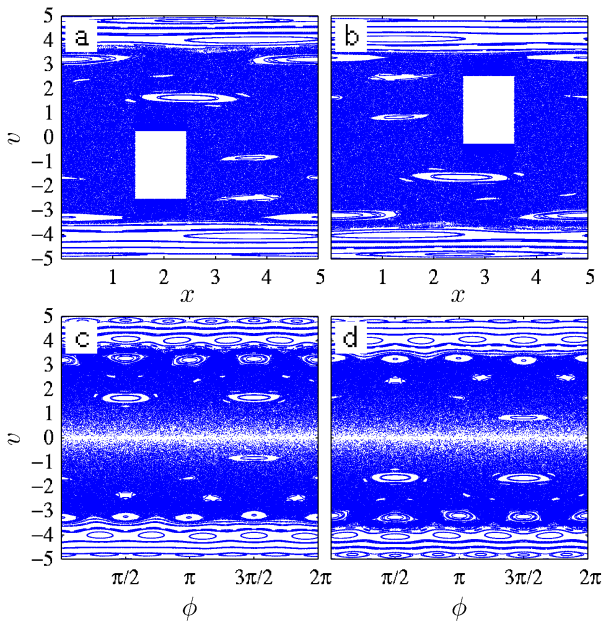}
\caption{\label{figure:sections_2} a) and c) PSS for a uniformly driven lattice with parameters $A\approx 0.57,\ \Delta \Phi=0$ and $\omega=1.0$.
		b) and d) PSS for $A\approx -0.57,\ \Delta \Phi=0$ and $\omega=1.0$. 
	     a) and b) phases and velocities are recorded at positions $x\ mod\ L=0$. c) and d) position $x$ taken modulo $L$ as well as the velocity at times $\omega t\ mod\ 2\pi=0$.}
\label{figure:sections}
\end{figure}

\begin{figure}[htbp]
\centering
\includegraphics{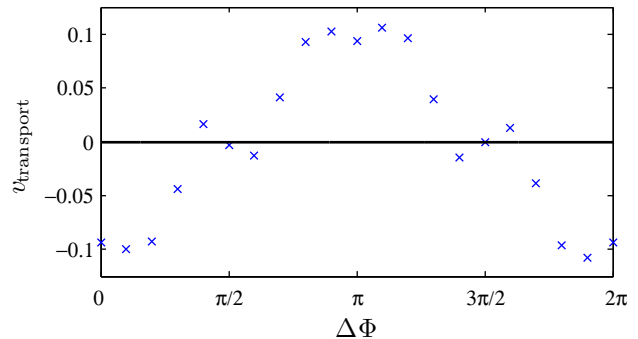}
\caption{Numerically determined transport velocity as a function of $\Delta \Phi$ in an infinite uniformly driven lattice with $A\approx0.57$ and $\omega=1.0$.}. 
\label{figure:transport}
\end{figure}

Even though the focus of this work is on composite systems consisting of multiple blocks exposed to different driving laws, these blocks are considered to be large in a sense that 
the dynamics of a particle within one block can be described by the Poincar\'{e} surfaces of section (PSS) as obtained 
by extending this block to an infinite uniformly driven lattice.
It is therefore sensible to discuss the dynamical features such as the transport properties as well as the appearance of the PSS of the uniformly driven lattice.
It is well established that the PSS for a uniformly driven lattice can be obtained by exploiting the temporal symmetry of the Hamiltonian: $H(x,t)=H(x,t+T)$. Thus an appropriate choice for 
the PSS is given by $M=\lbrace (x(t+kT)\ mod\ L,\ p(t+kT))\ |\ k \in \mathbb{N}\rbrace$ with $T=\frac{2\pi}{\omega}$ being the temporal period. 
According to this, we record the position taken modulo $L$ and the velocity at certain times $2\pi k,\ k \in \mathbb{N}$ (and call this 'position velocity section'). Such a PSS for 
$\omega=1.0$, $A\approx 0.57$ and $\Delta \Phi=0$ is shown in Fig. \ref{figure:sections} b) and features the typical mixed phase space \cite{LL}, i.e. 
KAM islands embedded in a chaotic sea which is bounded by the first invariant spanning curve (FISC). 
Note that the white rectangle is caused by 
adding the potential energy for particles which are within a barrier at times when position and velocity are recorded. 
This is done to avoid discontinuities in the PSS caused by the discontinuous potential $V(x,t)$ (cf. \cite{Petri:2010}).
For later usage we show additionally 
the position velocity section for parameters: $\omega=1.0$, $A\approx -0.57$ and $\Delta \Phi=0$ in Fig. \ref{figure:sections} a).
Note that Fig. \ref{figure:sections} a) (b)) shows the PSS of the left block (right block) of the corresponding two block system (see section \ref{section:4}).\\
A second possibility to illustrate the phase space of a uniformly driven lattice is given by the 'phase velocity section' which exploits the spatial symmetry of the Hamiltonian: $H(x,t)=H(x+L,t)$. 
To this end we record the phases and velocities at certain positions $iL$. Hence the PSS is obtained from the set of points $M=\lbrace (t(x+kL)\ mod\ T,\ p(x+kL))\ |\ k \in \mathbb{N}\rbrace$ and the resulting
plot is shown in Fig. \ref{figure:sections} d) (again, the PSS for parameters as in Fig. \ref{figure:sections} a) is shown additionally in Fig. \ref{figure:sections} c) 
for later usage). 
Apparently, it features qualitatively the same domains as the position velocity section (Fig \ref{figure:sections} b)), i.e. 
ballistic islands which are embedded in a bounded chaotic sea. However, in contrast to the previous case the chaotic sea appears to be non uniformly filled with trajectories. 
This seeming contradiction to ergodicity can be resolved easily: 
According to ergodicity the chaotic sea in Fig. \ref{figure:sections} b) can assumed to be filled with a uniform measure.   
Hence the number of particles $\Delta N$ that pass $x_i=iL$ and therefore contribute to
the phase velocity section per time $\Delta t$ and velocity interval $\Delta v$ is given by $\frac{\Delta N}{\Delta t}=\rho \frac{\Delta x}{\Delta t} \Delta v=\rho v \Delta v$, 
where $\Delta x$ denotes the distance a particle travels in time $\Delta t$ and 
$\rho$ is the number of particles per phase space interval.
Therefore, the number of particles passing $x_i$ per velocity interval is $\frac{\Delta N}{\Delta v}=\rho v \Delta t$ 
and hence proportional to $v$.\\    
We now comment on the transport properties within a single block.
For $\Delta \Phi_B\neq n\frac{\pi}{2}$ ($n \in \mathbb{Z}$) the biharmonic driving law breaks the time-reversal invariance as well as the parity symmetry of the Hamiltonian and 
the driven lattice allows for directed transport phenomena \cite{Flach:2000}.  
The transport as a function of $\Delta \Phi$ with fixed $A\approx 0.57$ and $\omega=1.0$ is determined numerically 
by simulating $10^5$ particles in a uniformly driven lattice for $10^6$ barrier oscillations and calculating their average velocity after a certain transient time. 
The results are shown in Fig. \ref{figure:transport} and appear to be in good agreement to the results of
a symmetry analysis (cf. \cite{Flach:2000, Quintero:2010}) which yields $v_{\text{transport}} \propto -\cos{(\Delta \Phi)}$. 
However, there are noticeable deviations, e.g. the reversed sign close to $\Delta \Phi_B=\pi/2$ and $\Delta \Phi_B=3\pi/2$. 
Note that these deviations should not surprise us because the authors in \cite{Flach:2000, Quintero:2010} considered continuous potentials instead of discontinuous potential barriers.     
For a more detailed analysis of the dynamics in a uniformly driven lattice, we refer to \cite{Petri:2010} where this was done in great detail.\\

%%%%%%%%%%%%%%%%%%%%%%%%%%%%%%%%%%%%%%%%%%%%%%%%%%%%%%%%%%%%%%%%%%%%%%%%%%%%%%%%%%%%%%%%%%%%%%%%%%%%%%%%%%%%%%%%%%%%%%%%%%%%%%%%%%%%%%%%%%%%%%%%%%%%%%%%%%%%%%%%%%%%%%%%%%%%%%%%%%%%%%%%%%%%%%%%%
\section{Interface conversion in the two block system}
\label{section:4}

In Ref. \cite{Petri:2011} it was argued that a BL as introduced in section \ref{section2} offers the opportunity for conversion processes from diffusive- to ballistic motion and vice versa at interface positions where the 
driving law changes.
These type of processes will be analysed in detail throughout this section. We demonstrate in particular their influence on the velocity distribution of a particle ensemble.\\

\subsection{Interface conversion}
\label{subsection:4.1}

Let us introduce the simplest possible finite BL which is build up of only two blocks (i.e. $N_{\text{Bl}}=2$) equipped with different driving laws. 
Such a system extends from $x_{\text{min}}=0$ to $x_{\text{max}}=2NL$ (so the simulation is stopped for a particle once it passes either of these positions)
and the driving laws are $d_0(t)$ for $x<x_{\text{mid}}=NL$ and $d_1(t)$ for $x\geq x_{\text{mid}}$.
The parameters of $d_1(t)$ are chosen as before ($\omega_1=1.0$, $A_1\approx 0.57$ and $\Delta \Phi_1=0$) and thus the dynamics within the 'right block' (RB) can be described by the two PSS in Fig. \ref{figure:sections} b) and d).
Moreover, Fig. \ref{figure:transport} reveals that the used driving law induces a negatively directed current. 
For the 'left block' (LB) we chose $\omega_0=1.0$, $A_0\approx-0.57$ and $\Delta \Phi_0=0$ yielding $d_0(t)=-d_1(t)$.
Hence the corresponding position-velocity section is given by Fig. \ref{figure:sections} a) and the phase-velocity section is the one shown in Fig. \ref{figure:sections} c).
The induced current in the LB is therefore of the same magnitude as in the RB, but positively directed.\\
To understand how this setup allows for conversion processes it is helpful to consider the dynamics of a particle with initial conditions in the chaotic sea of the LBs phase space. 
Due to the positively directed current, this particle is in the average transported towards $x_{\text{mid}}$
and the chaotic sea for positive velocities in the corresponding phase velocity section (Fig. \ref{figure:sections} c)) marks all possible phase space coordinates $(v_D,\phi_D)$ at which the particle can reach $x_{\text{mid}}$ diffusively,
while the coordinates belonging to ballistic motion (i.e. within ballistic islands or regular spanning curves above the FISC) $(v_B,\phi_B)$ are prohibited.
However, once the particle passes the interface at $x_{\text{mid}}$ its dynamics is no longer governed by the LBs phase space, but by the phase space of the RB, which is appropriately described by the PSS in Fig. \ref{figure:sections} d). 
The crucial observation is that some of the coordinates $(v_D,\phi_D)$ belonging to diffusive motion in the LB correspond to ballistic motion in the RB. This is best seen by means of a concrete example:
Imagine the particle reaches $x_{\text{mid}}$ with $(v=0.8,\phi=3\pi /2)$, which is inside the chaotic sea of the LBs PSS (Fig. \ref{figure:sections} c)). For $x>x_{\text{mid}}$ the particles 
dynamics is described by the RBs PSS (Fig. \ref{figure:sections} d)) where these coordinates correspond to a ballistic island. Hence, this initially diffusive particle would have become ballistic at the interface and we 
refer to this process as diffusive to ballistic motion conversion. Besides being injected into ballistic islands, the particles can equally well be injected into regular curves above the RBs FISC, because the FISC 
for positive velocities in the LB is at higher velocities as it is in the RB. To state a general rule, initially diffusive particles can be injected into every regular structure of the RBs PSS which has at least some 
'overlap' with the chaotic sea of the LBs PSS. An example of a regular structure in which no injection can occur is the chain of ballistic islands at $v\approx3.2$ in the RBs PSS (Fig. \ref{figure:sections} d)). These islands are 
'covered' by a chain of larger islands at the same velocity in the LBs PSS (Fig. \ref{figure:sections} c).

\subsection{Density evolution in the two block setup}
\label{subsection:4.2}

\begin{figure}[htbp]
\centering
\includegraphics{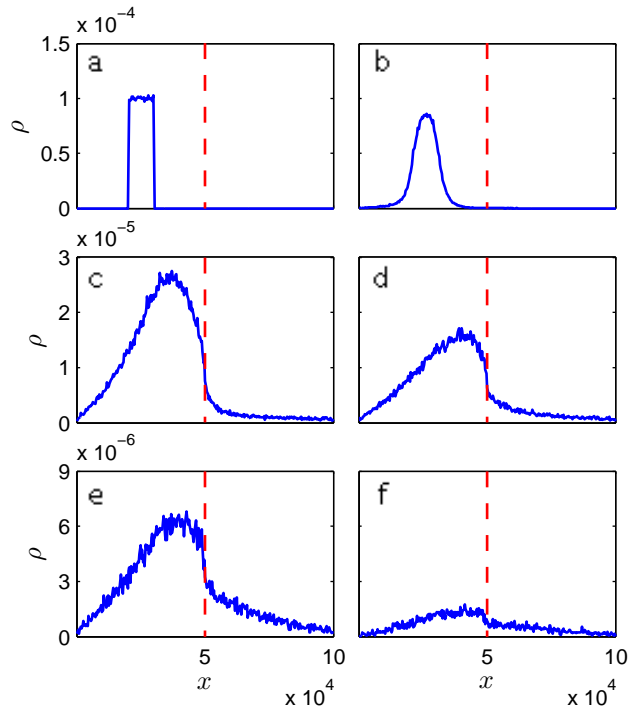}
\caption{Particle density at different times in the two block setup ($N_{\text{Bl}}=2$) and $N=10^4$ yielding $x_{\text{mid}}=5\cdot 10^4$ (indicated by the red dashed line) and $x_{\text{max}}=10^5$.
	    The two driving laws are $d_0(t)$ with $\omega_0=1.0$, $A_0\approx-0.57$ and $\Delta \Phi_0=0$ for $x<x_{\text{mid}}$ and $d_1(t)=-d_0(t)$ for $x>x_{\text{mid}}$.  
	    All densities are normalised to the initial particle number $10^5$. Snapshots of the particle density are shown for $t=0$ (a), $10^4$ (b), $10^5$ (c), $1.8 \cdot 10^5$ (d), $3\cdot 10^5$ (e) and $5.5\cdot 10^5$ (f).} 
\label{density}
\end{figure}

After having discussed the process of interface conversion in BLs, we will explore their influence on the time evolution of the particle density in the following.\\
To this end we propagate the dynamics of a particle ensemble in the two block system, which we introduced in section \ref{subsection:4.1}. As initial conditions 
we chose uniform distributions for the particles positions as well as their velocities with $0.4NL<x_{\text{ini}}<0.6NL$ and $-0.1<v_{\text{ini}}<0.1$ respectively.  
Hence the particles are symmetrically distributed around the LBs center and started in the chaotic sea of the phase space.
Naively one might expect that due to the oppositely directed currents in the LB and the RB an accumulation of particles might happen at the interface at $x_{\text{mid}}$. 
As we shall see in the following, this does not occur due to the previously introduced conversion processes.\\
Fig. \ref{density} shows snapshots of the normalised particle density at different times.
Fig. \ref{density} a) shows the particle density $\rho(x)$ for $t=0$. For $t>0$ the particle distribution starts to spread and reaches a Gaussian like shape at $t=10^4$ (Fig. \ref{density} b)). 
Afterwards the ensemble drifts in positive $x$-direction and once a sufficient amount of particles arrives at $x_{\text{mid}}=5\cdot10^4$ (marked by the red dashed line in Fig. \ref{density})
a sharp decrease of the density emerges at this position 
(Figs. \ref{density} c) and d)). This effect outlasts until $t\approx5.5\cdot 10^5$ (Figs. \ref{density} e) and f)). 
Finally at $t\approx 10^6$, all particles 
have left the system at either $x=x_{\text{min}}$ or $x=x_{\text{max}}$ and thus the density in the system is zero.\\
The broadening of the peak within the LB can be explained by underlying diffusion processes, because all particles are initially restricted to the chaotic sea. 
In fact, as long as the particles have not reached $x=x_{\text{mid}}$ 
the ensemble is super-diffusive \cite{Petri:2010}, leading to a fast expansion. 
On the one hand, the observed average drift of the ensemble in the LB is explained easily by the positively directed transport in the LB. On the other hand, the fast density decrease 
is -as mentioned before- somewhat counterintuitive. However, this effect is a straightforward consequence of the in section \ref{subsection:4.1} introduced conversion processes at $x_{\text{mid}}$. 
According to our discussion, the initially diffusive particles can be injected from the chaotic sea of the LBs PSS into regular structures of the RBs PSS which leads to a fast ballistic motion away from $x_{\text{mid}}$.
If the particle remains 
diffusive, the directed transport brings it back to the interface and an injection from the chaotic sea of the RB to regular structures of the LB is possible.  
Hence, this process is repeated until 
an injection occurs and the particle leaves the system at $x_{\text{max}}$ (or $x_{\text{min}}$) within a regular structure of the RB (or the LB).
The fact that we do not observe a particle accumulation at the interface demonstrates that the conversion process from diffusive to ballistic motion happens on a sufficiently fast timescale to 
overcompensate accumulation effects caused by the directed currents.

\subsection{Analysis of escaping particles in the two block setup}
\label{subsection:4.3}

In the present section we investigate the conversion processes from diffusive to ballistic motions further and  
illustrate an intriguing hallmark of these processes, namely their influence on the particles phase-velocity distribution.\\
The starting point is the same two block setup as before and the initial conditions are chosen as before, too. 
However, instead of discussing the particle positions at certain times as we did in section \ref{subsection:4.2}, we now record the particle velocities and phases at distinguished positions: $x_{\text{min}}$ and $x_{\text{max}}$.
\begin{figure}[htbp]
\centering
\includegraphics{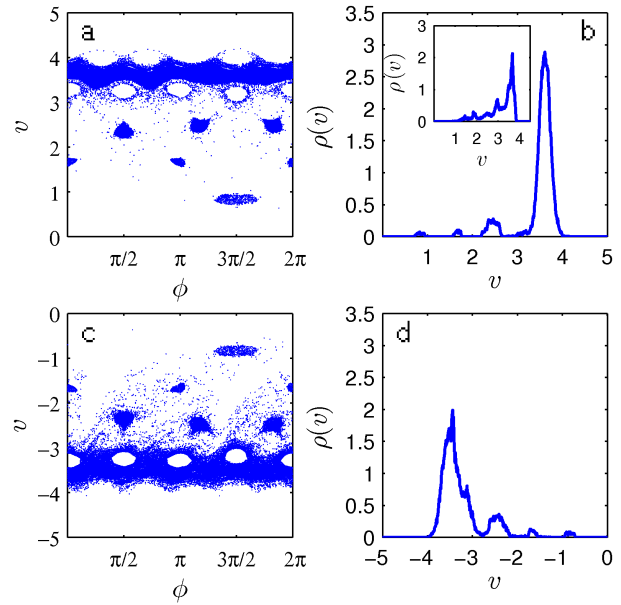}
\caption{\label{figure:phaseplots} Properties of the escaping particles for the two block setup with the same parameters as in Fig.\ref{density}.
a) Phases and velocities at $x=x_{\text{max}}$, c) at $x=x_{\text{min}}$.
	  b) and d) show the corresponding normalised velocity distributions $\rho(v)$. The inset in b) shows the velocity distribution for particles which arrive at $x=x_{\text{mid}}$ for their first time.}
\end{figure}  
To this end 
the phase $\phi$ and the velocity $v$ for every particle at $x_{\text{max}}$ are recorded and the result is shown in Fig. \ref{figure:phaseplots} a).
In the low velocity
regime ($v\lesssim3.0$) a distinguished island structure is apparent,
whereas for higher velocities ($v\gtrsim 3.5$) the particles possess all possible phases from $\phi=0$ to $\phi=2\pi$. 
In between ($3.0\lesssim v \lesssim3.5$) 
the particles appear to have randomly distributed phases and velocities, but do not occupy certain islands.\\
\begin{figure}[htbp]
\centering
\includegraphics{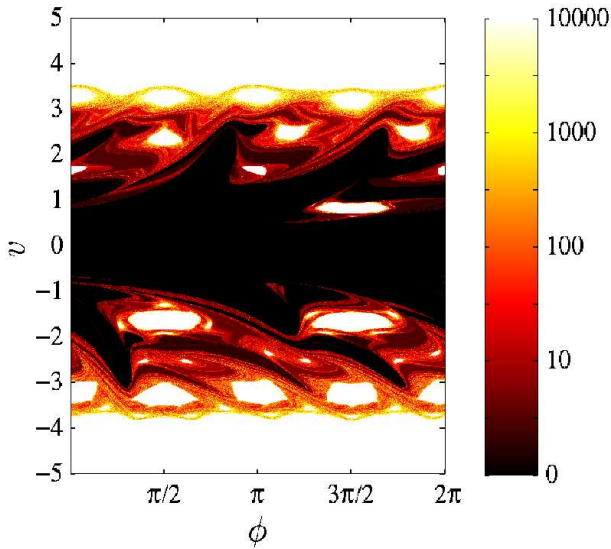}
\caption{\label{figure:levi}Length of L\'{e}vy flights for particles exhibiting chaotic dynamics in a uniformly driven lattice with driving law $d_1(t)$ (parameters as in Fig. \ref{figure:sections} d)). 
 The white regions correspond to initial conditions leading to regular motion.} 
\end{figure} 
The islands in the low velocity regime are evidently a consequence of
the diffusive to ballistic motion conversion processes. Once a particle which comes from the LB is injected at $x=x_{\text{mid}}$ into a regular structure of the RB, it cannot become diffusive again and travels to 
$x=x_{\text{max}}$ ballistically. Moreover, it is unlikely for a diffusive particle to reach $x_{\text{max}}$, and indeed impossible if the length of the block 
tends to infinity, because the local current in the RB is negative. Hence almost every particle in the velocity regime
$v\lesssim3.0$ in Fig. \ref{figure:phaseplots} a) is a ballistic one. 
Note that the island structures can easily be identified with the ballistic islands in the RBs PSS (Fig. \ref{figure:sections} d)). 
In an analogous way we can understand the velocity regime $v\gtrsim 3.5$: These particles are injected into surface spanning curves above the RBs FISC at $x_{\text{mid}}$. Accordingly, they are not restricted to 
certain phases. \\
The explanation why certain islands in the velocity regime $3.0\lesssim v \lesssim3.5$ are avoided by particles at 
$x_{\text{max}}$ can be given straightforwardly after our previous discussions on the conversion process. As already mentioned in section \ref{subsection:4.1}, these islands correspond 
to regular structures of the PSS in the RB which have no overlap with the chaotic sea of the LBs phase space and hence no injection occurs. \\
Finally, we turn our focus on the spreaded particles with $3.0\lesssim v \lesssim3.5$: A comparison with the corresponding PSS (Fig \ref{figure:sections} d)) reveals that these particles are located within the 
chaotic sea. Thus they have indeed passed the RB contrariwise to the directed current. To understand why this occurs predominantly in this velocity regime, a short detour on the 
typical length of L\'{e}vy flights in the driven lattice is necessary.
To this end we have simulated particles in a uniformly driven lattice with driving law $d_1(t)$ starting at $x=0$ with $2\cdot 10^6$ different initial conditions
covering the phase space interval $(0\le \phi_{\text{ini}}\le 2\pi,\ -5\le v_{\text{ini}} \le 5)$.
For every initial condition the number of barriers that the particle passes, before the sign of its velocity changes, is recorded and
shown in Fig. \ref{figure:levi} (initial conditions leading to regular motion were excluded and are shown in white).      
Apparently, the number of passed barriers before the velocity is reversed can differ by several orders of magnitude and strongly depends on the initial condition. 
Most interesting for our purpose is the observation that particles started in the velocity regime $3.0\lesssim v \lesssim3.5$ exhibit extraordinary long 'ballistic like' flights, which can be at the order of a few thousand barriers.
Hence it is more likely for a particle -that remains diffusive once it passes $x_{\text{mid}}$- to reach  
$x_{\text{max}}$ before being transported back to $x_{\text{mid}}$ if it is within this region of extraordinary long L\'{e}vy flights.
Indeed, a comparison of Fig. \ref{figure:levi} with Fig. \ref{figure:phaseplots} a) reveals, that the regions of long L\'{e}vy flights coincide with the ones where diffusive particles reach $x_{\text{max}}$.\\   
Even though the overall appearance of the plot shown in Fig. \ref{figure:levi} does strongly depend on the used parameters in the driving law, it is -for later usage- worth emphasising that 
the tendency for fast particles to exhibit much longer L\'{e}vy flights than slower ones is a rather general feature in the driven lattice. This is mainly caused by two facts:
Firstly, the lattice becomes a smaller perturbation for faster particles. Hence the average velocity change at a collision with a barrier is small for a particle which is close to the FISC (note that it is indeed zero for particles on the FISC) 
and as a consequence it takes many collisions with the barrier before a notable impact on the particle velocity occurs. Secondly, for particles with large initial velocities it is likely to become sticky to the FISC 
which -according to the discussions in \cite{Petri:2010}- leads to long ballistic flights.\\
Besides leaving the system at $x_{\text{max}}$, there is also the possibility for a particle to leave the system at $x_{\text{min}}$. Again, phase and velocity at this particular position are 
recorded and shown in Fig. \ref {figure:phaseplots} c).
This plot features qualitatively the same occupied domains as the one for $x_{\text{max}}$ but now mirrored at $v=0$: Distinguished island structures belonging to a regular ballistic dynamics for less negative velocities,
particles obeying chaotic dynamics which avoid ballistic islands for $-3.5\lesssim v \lesssim-3.0$ 
and particles on regular spanning curves with velocities $v \lesssim -3.5$. The main difference appears to be the larger amount of spreaded diffusive particles for velocities with $v\gtrsim -3$ which do not
correspond to ballistic islands. \\
Both the difference as well as the similarity to the $x_{\text{max}}$ plot can be understood intuitively: Since the ensemble is initially located around the center of the LB, the number of barriers 
that these particles have to pass diffusively and opposite to the direction of the local current is roughly $5000$, while it would be $10000$ for particles that have reached $x_{\text{mid}}$. 
Consequently, most of the spread diffusive particles seen 
in Fig. \ref {figure:phaseplots} c) are particles which have never reached $x_{\text{mid}}$. In contrast to this, the particles reaching $x_{\text{min}}$ within either ballistic islands or regular spanning curves have obviously 
passed $x_{\text{mid}}$ at least twice, because otherwise they could not be injected into the corresponding regular structures of the LB. Note that once a particle reaches $x_{\text{mid}}$, 
it is injected into the PSS of the RB with $d_1(t)$ for positive velocities and into the PSS of the LB with $d_0(t)$ for negative velocities. 
Since we chose $d_0(t)=-d_1(t)$, the ballistic islands apparent in Fig. \ref {figure:phaseplots} c) are the same as in Fig. \ref {figure:phaseplots} a), but mirrored at $v=0$.\\

\subsection{Injection probabilities into different regular structures}
\label{subsection:4.4}

So far we have seen that the two block system allows for diffusive to ballistic motion conversion processes. 
In this section we further investigate this phenomenon and discuss how likely injections into different regular regimes such as ballistic islands or spanning curves above the FISC are.\\
To get some insight it is instructive to compare the normalised phase integrated velocity distributions at $x_{\text{max}}$ (Fig. \ref {figure:phaseplots} b)) with the one at $x_{\text{min}}$ (Fig. \ref {figure:phaseplots} d)).
The for our purpose crucial observation is that 
the peak at high velocities ($\left| v \right| \gtrsim 3$) is less pronounced at $x_{\text{min}}$, and that the peaks corresponding to the ballistic islands appear slightly stronger populated (hardly visible) 
compared to the peaks in the distribution at $x_{\text{max}}$.
\begin{figure}[htbp]
\centering
\includegraphics{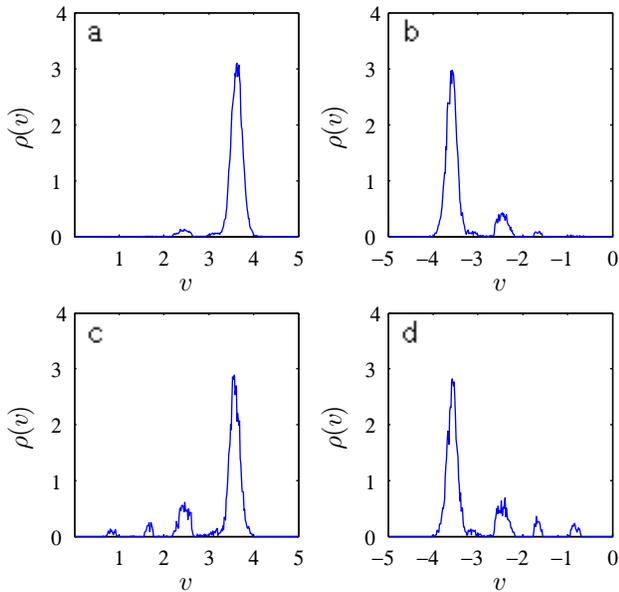}
\caption{\label{velos_cross}Normalised and phase integrated velocity distributions at $x_{\text{max}}$ in a) and c) and at $x_{\text{min}}$ in b) and d) for the 
subset of the ensemble which has crossed $x_{\text{mid}}$ $n_{\text{cr}}$ times (a) $n_{\text{cr}}=1$, b) $n_{\text{cr}}=2$, c) $n_{\text{cr}}=3$ and d) $n_{\text{cr}}=4$). Parameters as in Fig. \ref{density}.}
\end{figure}  
This effect can partially be explained
by the larger number of diffusive particles at $x_{\text{min}}$, leading to a broadening of the high velocity peak, but is also caused by certain characteristics of the injection process at $x_{\text{mid}}$
as we shall explain in the following.\\
A further understanding can be obtained by considering the velocity distributions for outgoing particles for a given number of 
times a particle has crossed $x_{\text{mid}}$ before it leaves the system 
(in the following we refer to this number as $n_{\text{cr}}$) as shown in Fig. \ref{velos_cross}.  
Apparently, for $n_{\text{cr}}=1$ (Fig. \ref{velos_cross} a))
the distribution features a very pronounced peak at velocities between 3 and 4, which appears stronger populated compared to the peak in the total velocity distribution at $x_{\text{max}}$ shown in Fig. \ref{figure:phaseplots} b).
On the contrary, the peaks corresponding to ballistic islands at smaller velocities are clearly weaker pronounced.  In the $n_{\text{cr}}=2$ case it is less apparent, but the islands at small velocities
are still less pronounced as for the total distribution in Fig. \ref{figure:phaseplots} d). For $n_{\text{cr}}=3$ this effect is reversed and the low velocity peaks contain relatively more particles than they do in
the total velocity distribution. Finally, particles leaving the system with $n_{\text{cr}}=4$ are very similar to the ones with $n_{\text{cr}}=3$ and no obvious deviation in the velocity
distributions is observed. It will become clear later, that this behaviour is strongly correlated with a different effect which is
worth mentioning at this point:  
Not only the normalised distributions differ for different $n_{\text{cr}}$, but the probabilities for a particle to leave the system after a certain number of crossing $n_{\text{cr}}$ are different as well.
Hence, the injection probabilities into any kind of regular structures $p_{n_{\text{cr}}}$ has to be different for different $n_{\text{cr}}$ and we indeed found 
numerically, that the probability for a particle to be injected
into a regular structure while it passes $x_{\text{mid}}$ for the first time is $p_1 \approx 0.49$, while it is $p_2 \approx 0.24$ for the second passing of $x_{\text{mid}}$ and 
approximately $p_i \approx 0.15$ for $n_{\text{cr}}>2$.\\
To understand both the different overall injection probabilities $p_{n_{\text{cr}}}$ for different $n_{\text{cr}}$ as well as the different appearances of the corresponding 
velocity distributions, an argument which combines the length of L\'{e}vy flights in different regions of the phase space (as discussed before and shown in Fig. \ref{figure:levi} for the RB and to be mirrored at $v=0$ 
for the LB)  
together with the ergodicity property is required: 
Due to ergodicity parts of the phase space corresponding to long ballistic flights must be visited less frequently, but ones a particle gets there, it stays for a comparably long time.
Hence, a particle initially started very close to 
$x_{\text{mid}}$ with a small velocity, reaches the interface after only a few collisions and as a consequence it is unlikely to reach the high velocity regime for such a particle. On the contrary, for a particle that started 
far -say some thousand barriers- away from $x_{\text{mid}}$, it is likely that the particle reaches this high velocity domain at some point. Once it possesses such a high velocity, the length of its L\'{e}vy flight 
is of the same order as its distance from $x_{\text{mid}}$
and the particle typically reaches the interface while being still confined to this domain of phase space.
We remark that this effect can easily be observed in the velocity distribution at $x_{\text{mid}}$ for particles which reach this position for their first time as shown in the inset of Fig. \ref{figure:phaseplots} b).
Apparently the distribution reveals a strongly pronounced peak at velocities close the LBs FISC in agreement with the previous discussion.
Since these fast particles in the LB have velocities $v\gtrsim 3.5$ they are injected into regular spanning curves above the FISC of the RB, 
which explains why particles at the first injection process have extraordinary high velocities and consequently why $p_1$ is extraordinary large.
Additionally, it illustrates why this domain appears to be strongly populated for $n_{\text{cr}}=1$ (cf. Fig. \ref{velos_cross} a)). \\
Following the same arguments, the velocity distribution for $n_{\text{cr}}=2$ as well as the slightly enhanced value of $p_2$ compared to $p_i$ with $i>2$ can be understood too. 
Because the particles are extraordinary fast at their first arrival at $x_{\text{mid}}$, a comparably large fraction of particles which remain diffusive is injected into the part of the RBs PSS corresponding to 
long L\'{e}vy flights (cf. Fig. \ref{figure:levi}). Hence they surpass a large number of barriers in the RB  which is at the order of $10^3$ 
before their velocity is reversed for the first time. Afterwards, they are most likely transported towards $x_{\text{mid}}$ due to 
the negatively directed current in the RB.       
Now the same argument as before holds: because these particles are relatively far away from $x_{\text{mid}}$, they reach
$x_{\text{mid}}$ predominantly with a high velocity. Hence, this domain is still strongly populated for the $n_{\text{cr}}=2$ injection process and $p_2$ is slightly enhanced. 
By now most of the fast particle have left the system and the still 
diffusive ones are transported back to $x_{\text{mid}}$ after only a few collisions. Consequently, the injection probability into the high velocity domain is suppressed for larger $n_{\text{cr}}$.\\

%%%%%%%%%%%%%%%%%%%%%%%%%%%%%%%%%%%%%%%%%%%%%%%%%%%%%%%%%%%%%%%%%%%%%%%%%%%%%%%%%%%%%%%%%%%%%%%%%%%%%%%%%%%%%%%%%%%%%%%%%%%%%%%%%%%%%%%%%%%%%%%%%%%%%%%%%%%%%%%%%%%%%%%%%%%%%%%%%%%%%%%%%%%%%%%%%
\section{Controlling the velocity distribution}
\label{section:5}

We have seen in section \ref{section:4} that a BL consisting of differently driven blocks allows for the conversion of diffusive- to ballistic motion and vice versa at interfaces where the driving law changes.
As a hallmark of these processes, we obtained velocity distributions with pronounced peaks at the velocities associated to regular structures in the underlying phase space. In the following section
we investigate to what extend these conversion processes can be exploited to modulate the velocity distribution for outgoing particles in a controlled manner by adjusting parameters in the driving law.\\
Again we focus on the two block setup ($N_{\text{Bl}}=2$, $N=10^4$) with two different driving laws $d_0(t)$ for $x<x_{\text{mid}}$ and $d_1(t)$ for $x\geq x_{\text{mid}}$. 
The parameters in $d_0(t)$ remain as before throughout the entire section and thus a positively directed current is induced in the LB.
On the contrary, each parameter in $d_1(t)$ is varied separately and its influence on the phase velocity distributions at $x_{\text{max}}$ and $x_{\text{min}}$ shall be studied. 
To this end we simulate an ensemble of $10^5$ particles (initial conditions as in section \ref{section:4}) until all particles left the two block system.\\
However, before we begin a detailed analysis of the phase velocity distributions for outgoing particles at $x_{\text{max}}$ in
various parameter regimes it is sensible to have a brief discussion on their expected parameter dependence: 
Assuming a negatively directed current in the RB (which will be the case in most of the studied scenarios), we can expect that the particles arrive at $x_{\text{max}}$ predominantly within 
either ballistic islands or regular curves above the FISC corresponding to the RBs phase space. Hence the phase velocity distributions at $x_{\text{max}}$ provide an 'image' of 
all the regular structures in the PSS of the RB which have overlap with the chaotic sea of the LBs PSS (Fig. \ref{figure:sections} c)). Additionally to the ballistic particles 
some diffusive particles are expected to reach $x_{\text{max}}$ if the underlying phase space of the RB possesses regions of long L\'{e}vy flights.
Provided with these arguments we are able to explain most of the phenomena occurring in the following sections.

\subsection{Frequency variations}
\label{section:5.1}

To begin with, we explore how a change of $\omega_1$ manifests itself in the particles' velocity distributions when they leave the system
at either $x_{\text{max}}$ or $x_{\text{min}}$. The other parameters of $d_1(t)$ are kept constant at $A_1\approx 0.57$ and $\Delta \Phi_1=0$ (inducing negatively directed currents in the RB for all values of $\omega_1$). \\
\begin{figure}[htbp]
\centering
\includegraphics{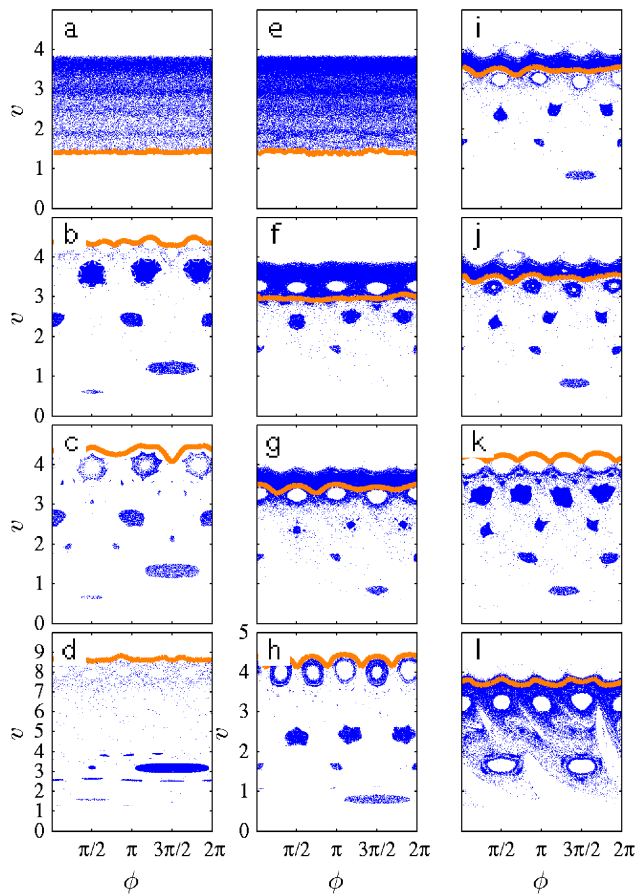}
\caption{\label{phaseplots_omega} Phase velocity distributions at $x=x_{\text{max}}$ for a two block system with $d_0(t)$ as in Fig. \ref{density}. The orange line indicates the FISC in the PSS of the RB.
  For $d_1(t)$ the parameters are $A_1\approx 0.57$, $\Delta \Phi_1=0$ and $\omega_1=0.05,\ 1.50,\ 1.65,\ 4.00$ for a)-d).
  For e)-h) we have $\Delta \Phi_1=0$, $\omega_1=1.0$, and $A_1=0.02,\ 0.40,\ 0.50,\ 1.00$.
 For i)-l) we have $A_1 \approx 0.57$, $\omega_1=1.0$, and $\Delta \Phi_1=0.1\pi,\ 0.2\pi,\ 0.5\pi,\ 1.0\pi$.} 
\end{figure}  
Figs. \ref{phaseplots_omega} a)-d) show the phase velocity distribution at $x_{\text{max}}$ with varying frequencies $\omega_1=0.05,\ 1.50,\ 1.65$ and $4.00$. 
Fig. \ref{phaseplots_omega} a) corresponds to the low frequency ($\omega_1=0.05$) regime where the barriers within the RB move very slowly compared to the barriers inside the LB. 
It shows a broad velocity band which is sharply confined  
between $v\approx 1.4$ and $v\approx 3.8$ where particles reach $x_{\text{max}}$ without any
apparent restrictions on their phases. 
This can be explained straightforwardly, by considering
the expansion of the chaotic sea. As argued in \cite{Petri:2010}, the velocity for particles on the first invariant spanning curve (FISC) limits the chaotic sea and can be estimated by 
$v_{\pm}=\pm \ \dot d_{max}^{\pm}\ \pm \ \sqrt{2V_0}$ where $\pm$ correspond to positive or negative velocities and $\dot d_{max}^{\pm}$ is the maximal barrier velocity 
in either positive- or negative direction.   
For a small frequency $\omega_1$ the maximal velocity of the barrier in any given direction becomes small as well, leading to a FISC in the RB at very low      
velocities (for illustration we show the FISC in the RBs PSS as an orange line in Fig.\ref{phaseplots_omega}). 
According to our previous discussion, almost all particles are injected into regular spanning curves above the RBs FISC at $x_{\text{mid}}$ and are able to reach $x_{\text{max}}$ at arbitrary phases.
The sharp cutoff at $v\approx 1.4$ can be explained by employing the approximation of a static potential which yields a minimal particle velocity of $v=\sqrt{2V_0}\approx1.41$ 
to surpass the barrier.\\ 
Increasing $\omega_1$ in the RB leads us to a second regime, where the FISCs in both blocks are at similar velocities
and we show the corresponding phase velocity distribution for $\omega_1=1.50$ in Fig. \ref{phaseplots_omega} b).
The distribution reveals the expected domains of particles in ballistic islands and some diffusive particles close to the FISC in the RBs phase space. Both features are 
straightforward consequences of the previous discussions and we refrain from a reiteration of the arguments. However it is appealing to explore how the overall picture changes for 
a small variation in $\omega_1$.
To this end the phase velocity plot
for another frequency of $\omega_1=1.65$ is presented in Fig. \ref{phaseplots_omega} c). Apparently, the phase velocity distribution of particles at $x_{\text{max}}$ looks similar 
to the one for $\omega_1=1.50$. 
Nevertheless there are some differences worth emphasising. First of all, the 'large' islands survive the frequency change, but are shifted to slightly higher velocities for an increased $\omega_1$.
Additionally, the islands at $v\approx4.0$ appear now emptied
around their center.   
\begin{figure}[htbp]
\centering
\includegraphics{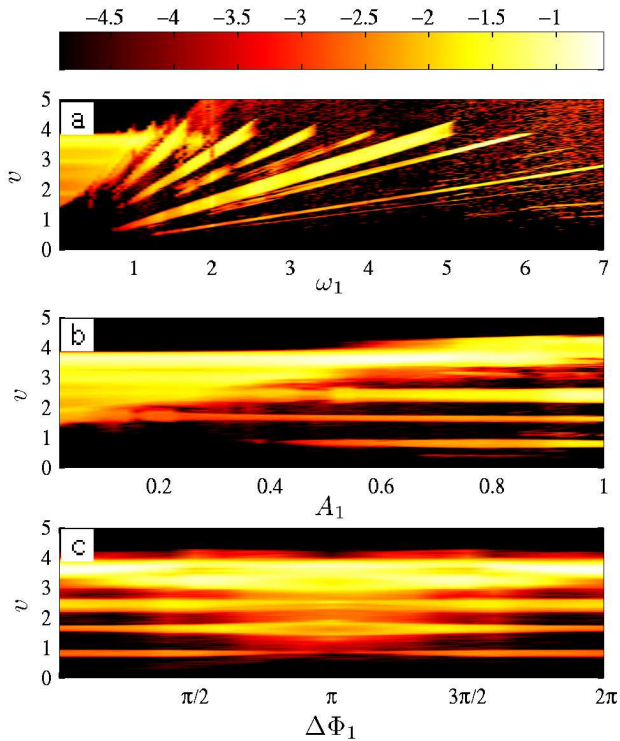}
\caption{\label{colourplots}Phase-integrated velocity distributions at $x_{\text{max}}$ (setup as as in Fig. \ref{density})
on a logarithmic scale for variations of a) $\omega_1$, b) $A_1$ and c) $\Delta \Phi_1$. Remaining parameters as in Fig. \ref{phaseplots_omega}.} 
\end{figure}  
By comparing the PSS of the LB and the RB we see that those depleted parts of the ballistic islands
in the RBs phase space are at velocities which are partially above the FISC of the LB.  
Since the injection at $x_{\text{mid}}$ can only occur for regular curves within the island that are at least in parts covered by the chaotic sea of the LBs phase space and the outer curves in each island reach to
lower velocities than the inner curves, the latter ones are depopulated first.\\
Finally, we address the regime where the FISC in the RB is at much higher velocities than the FISC in the LB, which is realised for $\omega_1>>\omega_0$ and
the phase velocity distribution is shown exemplarily for
$\omega=4.0$ in Fig \ref{phaseplots_omega} d). 
Apparently, only the island initially at ($\phi \approx 3\pi/2,\ v\approx1.2$) survives,
but is shifted to considerably higher velocities ($v\approx 3.0$). Moreover, we notice a comparably large portion of diffusive particles at high velocities ($7.0\lesssim v\lesssim 9.0$) below the FISC in the RB.\\
To visualise the results over a broad range of $\omega_1$, we integrate the phase velocity distributions over the phase 
and normalise each in the range $\omega_1=0.05\ -\ 7.00$ (see Fig.\ref{colourplots} a)). 
\begin{figure}[htbp]
\centering
\includegraphics{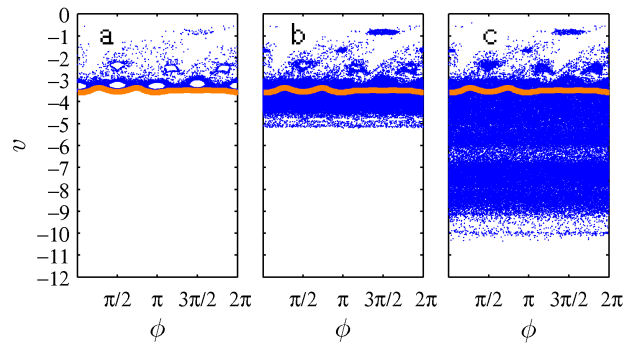}
\caption{\label{figure:omega_zero} Phase velocity distributions for particles which exit at $x_{\text{min}}$
 for a) $\omega_1=0.05$, b) $\omega_1=1.50$ and c) $\omega_1=4.00$ (remaining parameters as in Fig.\ref{density}). The orange line indicates the FISC position in the LB.} 
\end{figure}  
Evidently, the particles arrive within a broad range of velocities for small $\omega_1$'s, while
they are restricted to certain comparatively narrow velocity intervals at higher frequencies.
Both can be understood within the above analysis: For small frequencies the particles are predominantly injected into regular spanning
curves and for higher ones mainly into individual ballistic islands. These islands are shifted to higher velocities for increasing $\omega_1$ and once they pass the FISC of the LB
they simply disappear.
What these velocity distributions reveal additionally is that the islands mean velocities tend to increase linearly with $\omega_1$. 
This is caused by the fact, that ballistic islands correspond to trajectories synchronised with the barrier oscillations in a way that every collision with the barrier occurs at distinguished phases. 
Each island is thereby characterised by its winding number $n$ which is defined (within a block) as the number of unit cells the particle passes within one period $T$. Hence the average 
velocity of a particle trapped in such an island is given by $v_n=\frac{Ln}{T}=\frac{Ln}{2\pi}\cdot \omega_1$ and thus proportional to $\omega_1$. \\
\begin{figure}[htbp]
\centering
\includegraphics{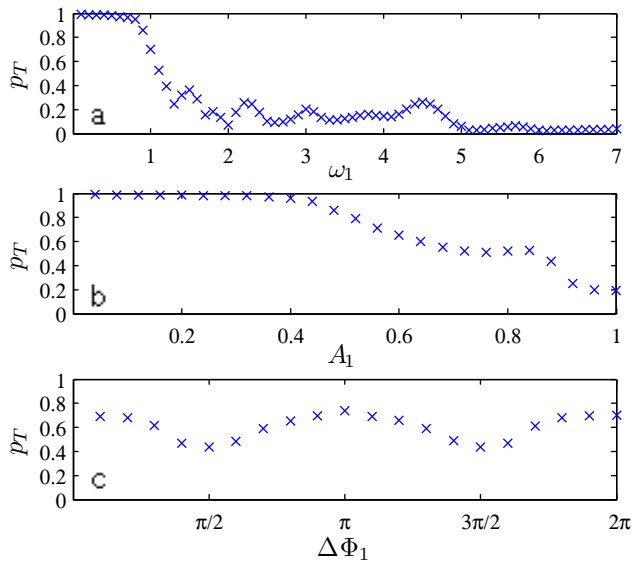}
\caption{\label{transmission_probs} Transmission probability $p_T$ as a function of a) $\omega_1$, b) $A_1$ and c) $\Delta \Phi_1$.
	  Remaining parameters as in Fig. \ref{phaseplots_omega}.} 
\end{figure} 
Besides the possibility for leaving the system at $x_{\text{max}}$ particles can also exit at $x_{\text{min}}$. Although, the parameters in the LB are kept constant, 
the parameters in $d_1(t)$ can have a substantial influence on the particle dynamics at $x_{\text{min}}$ due to conversion processes after multiple crossing of $x_{\text{mid}}$.
This is demonstrated in Fig. \ref{figure:omega_zero} where the phase velocity distributions are shown exemplarily for $\omega_1=0.05,\ 1.50$ and $4.00$.
Again let us discuss first the small frequency regime (Fig. \ref{figure:omega_zero} a)). Since the FISC in the RB is at very low velocities for $\omega_1=0.05$, all the fast particles 
are immediately injected into regular curves and therefore leave the system at $x=x_{\text{max}}$. Consequently, there is little probability for a particle to be re-injected into regular structures in the LB once it has entered the RB and 
as a consequence all the particles seen in Fig. \ref{figure:omega_zero} a) are diffusive ones. More precisely, these are mainly particles that never reached $x_{\text{mid}}$, i.e. the 'diffusive background' which is always present at $x_{\text{min}}$,
independent of the parameters of $d_1(t)$ and consists of approximately $20\%$ of all particles (note that this number tends to zero if the size of a block tends to infinity).\\
For increasing $\omega_1$ the situation changes substantially, as demonstrated 
for $\omega_1=1.50$ in Fig. \ref{figure:omega_zero} b). 
Evidently, some of the ballistic islands at minor negative velocities are filled. Additionally, we observe particles on regular spanning curves below 
the FISC, which were not present in the case $\omega_1=0.05$. 
However, before we give an explanation by means of the underlying PSS we remark that the relevant parts of the phase space are now the ones for negative velocities (cf. Figs. \ref{figure:sections} c) and d) for $v<0$). 
By keeping this in mind, the difference of Fig. \ref{figure:omega_zero} b) compared to Fig. \ref{figure:omega_zero} a) can be
understood intuitively by the FISC position: Since the FISC for negative velocities in the RB is now lower than the one for negative velocities in the LB, there is significant overlap
of the chaotic sea in the RB and ballistic islands- as well as invariant curves below the FISC in the LB. 
Consequently, these parts of phase space in the LB can be populated by diffusive particles in the chaotic sea of the RB which are transported back to $x_{\text{mid}}$ by the negatively directed current.\\
The same arguments hold for the high frequency domain that is exemplarily shown for $\omega_1=4.00$ in Fig. \ref{figure:omega_zero} c). 
Since the FISC for negative velocities in the RB is at lower, i.e. more negative, velocities compared to the case $\omega_1=1.50$ the particles which penetrate into the RB diffusively can now be injected 
into regular curves of the LB at more negative velocities 
and therefore reach $x_{\text{mid}}$ faster than before.
Note that even though the particles have no apparent modulation in phase, they are not uniformly distributed within the accessible range of velocities, which can be traced back to the underlying PSSs. 
For example the local minimum at $v\approx -6.5$ (cf. Fig. \ref{figure:omega_zero} c)) is caused
by a chain of ballistic islands embedded in the RBs chaotic sea which is decreasing the overlap and therefore the injection probability into regular structures of the LB within this velocity regime.\\
We conclude this chapter with some remarks on the impact of a varying frequency in the RB on the transmission probability $p_T$ through the interface. 
i.e. the ratio of the particles reaching $x_{\text{max}}$ and the total number of particles reaching $x_{\text{mid}}$.
Its dependence on $\omega_1$ is shown in Fig. \ref{transmission_probs} a) (note that we omit the diffusive background, i.e. we only take particles into account that have actually reached $x_{\text{mid}}$ at least once).
For small frequencies almost all particles are transmitted, while
this changes drastically for $\omega_1 \gtrsim 1.0$ when $p_T$ starts to oscillate around $p_T\approx 0.2$ until it is again decreased above $\omega_1 \approx 4.5$ to $p_T\approx 0.05$ where it appears to saturate.
All three regimes can be understood by means of our previous discussions: Since the particles arriving at $x_{\text{mid}}$ for their 
first time have predominantly high velocities ($3.5\lesssim v \lesssim4.0$) (cf. the inset of Fig. \ref{figure:phaseplots} c)),
$p_1$ (and hence $p_T$) is large whenever regular structures in the phase space of the RB are located within this velocity regime. For small $\omega_1$ this is the case because the PSS in the RB is in this velocity regime filled with 
regular curves above the corresponding FISC. For $\omega_1>1$ this is no longer true (cf. for example \ref{phaseplots_omega} b)) and thus the particles rely on injection processes into ballistic islands rather than regular spanning curves
to surpass the RB ballistically. Consequently $p_T$ drops significantly. For larger frequencies $p_1$ (and therefore $p_T$) increases whenever one of the ballistic islands in the RBs phase space is in the 
velocity regime $3.5\lesssim v \lesssim4.0$ (cf. Fig. \ref{colourplots} a)) and decreases once this island disappears because it is above the FISC of the LB, which explains the oscillatory behaviour of $p_T$ for $1.0<\omega_1<4.5$. 
At $\omega_1 \approx 4.5$ the last ballistic island
corresponding to an integer value of the winding number $n$ disappears and thus 
$p_T$ drops further.

\subsection{Amplitude variations}
\label{section:5.2}

Analogous to the previous discussion, we explore how a change of the oscillation amplitude $A_1$ in the RB affects the particle properties for fixed values of $\omega_1=1.0$ and $\Delta \Phi_1=0$.
As before the parameters in $d_0(t)$ remain unaltered. 
The corresponding phase velocity distributions at $x_{\text{max}}$ are shown exemplarily in Figs. \ref{phaseplots_omega} e)-h).
The regime of small amplitudes is represented by Fig. \ref{phaseplots_omega} e) which shows the particles 
phase velocity plot at $x_{\text{max}}$ for $A_1=0.02$. Apparently, the particles are restricted to the same velocity interval as for the 
'small frequency regime' illustrated in \ref{phaseplots_omega} a), namely $1.4\lesssim v \lesssim 3.8$. 
The interpretation in terms of a quasi static barrier motion as provided previously for the case of small frequencies holds also here for the case of a small amplitude motion.
Accordingly, most particles are injected into regular spanning curves above the FISC of the RB once they pass $x_{\text{mid}}$ for the first time.
If we increase $A_1$ the FISC is shifted to higher velocities and reaches $v\approx3.0$ for $A_1=0.40$. The associated phase velocity plot is shown in Fig. \ref{phaseplots_omega} f). 
For velocities below the FISC we observe the familiar island like structure. Particles can still be injected into curves above the FISC, where a chain of islands is avoided ($v\approx 3.2$).
The latter one is caused by the fact, that these ballistic islands in the phase space of the RB have no overlap with the chaotic sea in the LB. This is due a chain of larger ballistic islands 
in the LBs phase space (Fig. \ref{figure:sections} c)) at the same velocity.\\
Fig. \ref{phaseplots_omega} g) shows the corresponding graph for a slightly increased amplitude $A_1=0.50$ and reveals that
the FISC is -as expected- shifted to higher velocities. Furthermore, a region containing a considerable number of diffusive particles evolves at $3.0\lesssim v \lesssim 3.5$ followed by an island structure for even
lower velocities. The surprisingly high number of diffusive particles at velocities slightly below the FISC is caused by extraordinary long L\'{e}vy flights within this region due to a 
chain of cantori at $v\approx 3$. As a consequence of the small flux through this chain, the dwell time for particles in this part of the phase space is enhanced drastically \cite{Petri:2010}.
According to our previous discussions in section \ref{subsection:4.3} this enhances the probability for diffusive particles to reach $x_{\text{max}}$ in this part of the phase space.\\
Fig. \ref{phaseplots_omega} h) shows the phase velocity distribution at $x_{\text{max}}$ for a large amplitude $A_1=1.0$.  
Evidently, the FISC in the RB is in this case 
at higher velocities than it is in the LB. Thus, mainly particles within ballistic islands reach $x_{\text{max}}$. As we observed for a high frequency (cf. Fig. \ref{phaseplots_omega} c)), some of the islands are partially
above the LBs FISC and their inner curves are therefore depleted. 
The phase integrated results over a range of amplitudes $A_1=0.02-1.0$ is shown in Fig. \ref{colourplots} b).   
Similar to what we observe for the frequency dependence the particles cover a broad velocity interval ($1.5\lesssim v \lesssim4.0$) for small amplitudes. With increasing $A_1$ this interval decreases and additional narrow velocity
intervals emerge. 
The latter are due to ballistic islands in the PSS of the RB. 
We remark that opposite to the frequency dependence (Fig. \ref{colourplots} a)), the corresponding velocity peaks are barely affected by changes of the amplitude.
At least qualitatively, this behaviour can be understood by considering trajectories of particles within these ballistic islands: On the one hand the constant mean velocity of an island can be understood 
by remembering that ballistic islands correspond to synchronised orbits, where the particle collides with the barrier at distinguished phases. Hence, their mean velocity only depends on the frequency rather than on the
amplitude of the oscillation. On the other hand the precise point of appearance as well as the shape of the islands can very well depend on $A_1$ and has to be determined by numerical simulations.\\
The amplitude dependence of the transmission probability $p_T$ is shown in Fig. \ref{transmission_probs} b), where -as before- we consider only particles that reach $x_{\text{mid}}$ at least once. 
Evidently, $p_T$ remains approximately unity for $0\lesssim A_1 \lesssim0.4$ and thereafter decreases steadily with further increasing $A_1$, besides a weakly pronounced local maximum at $A\approx 0.8$. Following the arguments
presented in the previous section, this behaviour is a result of the overlap of the chaotic sea in the LB and regular structures
within the RB. \\
No relevant new phenomena are observed for the particles exiting at $x_{\text{min}}$ which is why we refrain from providing a discussion of this case.

\subsection{Phase variations}

Let us finally explore the impact of phase changes $\Delta \Phi_1$ for fixed $\omega_1=1.0$ and $A_1\approx 0.57$.  
The phase velocity distributions at $x_{\text{max}}$ are shown in Fig. \ref{phaseplots_omega} i)-l). 
Fig. \ref{phaseplots_omega} i) and j) correspond to comparably small phase differences of $\Delta \Phi_1=0.1\pi$ and $\Delta \Phi_1=0.2\pi$ respectively. A comparison of these graphs with the one for $\Delta \Phi_1=0$ (Fig. 
\ref{figure:phaseplots} a)) reveals that the main difference is a small shift of the entire phase space to the region of smaller phase values.
Consequently, the $n=4$ island ($v\approx 3.2$)
in the RB is not completely covered by the corresponding 
island in the LB anymore. Hence, the outer most curves can be populated by particles.
The tendency that islands are moved to smaller values of the phases for increasing $\Delta \Phi_1$ is still apparent for $\Delta \Phi_1=0.5\pi$ as seen in Fig. \ref{phaseplots_omega} k). Additionally,
the $n=4$ island is fully populated by particles and more diffusive particles manage to reach $x_{\text{max}}$. Finally, there are no particles in spanning curves above the FISC anymore.
In contrast to the frequency and amplitude dependence, $\Delta \Phi_1$ has a substantial influence on the symmetries of the Hamiltonian and thus 
on the transport (cf. Fig.\ref{figure:transport}). In fact for $\Delta \Phi_1=0.5\pi$ time reversal symmetry is restored ($d_1(t,\Delta \Phi_1=0.5\pi)=d_1(-t,\Delta \Phi_1=0.5\pi)$)
and the directed current vanishes.
Obviously, this increases the probability for a particle to traverse diffusively the RB,
which explains the notable amount of non ballistic particles in Fig. \ref{phaseplots_omega} k).
Finally, the transport in the RB is reversed, i.e. points in a positive direction, for $\Delta \Phi_1=\pi$. Moreover we notice that $d_1(t,\Delta \Phi=\pi)=-d_1(t+\pi,\Delta \Phi=0)$, i.e.
the driving law in the RB equals the driving law in the left one besides an initial phase shift of $\pi$. Consequently, the ballistic islands for $d_1(t)$ are at the same velocities but at phases shifted
by $\pi$. Since the PSS for $d_0(t)$ is 'almost' invariant under a 
shift of $\pi$, the overlap from the chaotic sea in the LB and regular structures in the right one is comparably small for $\Delta \Phi_1=\pi$. Thus, very few particles reach
$x_{\text{max}}$ ballistically and the phase velocity distribution is dominated by chaotic particles (Fig. \ref{phaseplots_omega} l)). \\
Fig. \ref{colourplots} c) shows the $\Delta \Phi_1$ dependence of the phase integrated velocity distributions. We observe an increasing amount of diffusive particles for $\Delta \Phi_1$ close to $\Delta \Phi_1=\pi$ compared to $\Delta \Phi_1=0$
(or $\Delta \Phi_1=2 \pi$). As stated before, this is a consequence of the dependence of the direction of the current in the RB which is positively directed for $\Delta \Phi_1=\pi$ 
while it is negatively directed for $\Delta \Phi_1=0$. \\
As a last remark on the two block setup the transmission probability $p_T$ as a function of $\Delta \Phi_1$ is shown in Fig. \ref{transmission_probs} c) and reveals an oscillatory behaviour with local maxima at $\Delta \Phi_1=0, \pi$ and $2\pi$.
For frequency and amplitude variations we argued that the value of $p_T$ is determined by the overlap of regular structures in the RB with the chaotic sea of the LB. However, 
in the present case, this overlap is minimal for $\Delta \Phi_1=\pi$. Apparently, this decreasing overlap for $\Delta \Phi_1 \rightarrow \pi$ is compensated by an increasing probability for a particle to surpass the RB diffusively due to 
the positively directed current (cf. Fig.\ref{figure:transport}).

%%%%%%%%%%%%%%%%%%%%%%%%%%%%%%%%%%%%%%%%%%%%%%%%%%%%%%%%%%%%%%%%%%%%%%%%%%%%%%%%%%%%%%%%%%%%%%%%%%%%%%%%%%%%%%%%%%%%%%%%%%%%%%%%%%%%%%%%%%%%%%%%%%%%%%%%%%%%%%%%%%%%%%%%%%%%%%%%%%%%%%%%%%%%%%%%%
\section{Velocity distributions in superlattices}
\label{section:6}

In the previous sections we have demonstrated how a setup build up out of two blocks with different driving laws allows for conversion processes from diffusive- to ballistic motion. 
Even more we were able to control the velocity distributions for outgoing particles at $x_{\text{max}}$ by adjusting parameters in the RB. In the following we 
argue how the so far gained insights can be exploited to maintain mono energetic- pulsed particle beams out of diffusive particle ensembles in superlattices containing a few hundred blocks.
The general outline of the used scheme is as follows: We start with an initially diffusive particle ensemble in the $B=0$ block (with $B$ being the block index, cf. Fig \ref{fig:lattice})
which is transported towards a first interface where particles can be injected into ballistic islands
of the $B=1$ blocks phase space.
The parameters in the driving laws are chosen such that these now ballistic particles travel opposite to the directed currents and thus we obtain a peaked velocity distribution at the end of the $B=1$ block.
For the following blocks with $B=2,...,100$ we show how the width 
of each peak in the velocity distribution can be tuned by adjusting the amplitude of the barrier oscillation blockewisely. As a last step, we demonstrate how an appropriate choice in the driving laws for $B>100$ allows to
preserve one of the peaks in the velocity distributions while the other peaks are 
subsequently removed. Thus we obtain a mono energetic particle beam for outgoing particles in the superlattice.  
Moreover, the beam is pulsed in a sense that the particles leave the systems only at distinguished phases.

%%%%%%%%%%%%%%%%%%%%%%%%%%%%%%%%%%%%%%%%%%%%%%%%%%%%%%%%%%%%%%%%%%%%%%%%%%%%%%%%%%%%%%%%%%%%%%
\subsection{Interface dynamics of ballistic particles}
\label{section:6.1}
 
Before we start a detailed discussion of the physics in the BL containing a few hundred blocks, let us again consider the simple case of a two block setup to introduce a new type 
of conversion processes which occurs in larger BL namely ballistic to ballistic- or ballistic to diffusive conversion.
To make our discussion more concrete we consider again a setup with driving laws as in section \ref{section:4}. Hence the PSS for the LB is shown in Fig. \ref{figure:sections} c) and  
the PSS for the RB is shown in Fig. \ref{figure:sections} d). In contrast to the previous discussions we explore the possible conversion processes for a ballistic particle arriving 
at the interface. For example, consider a particle beam started at $x_{\text{min}}$ which uniformly occupies the ballistic island at $(v=1.8,\phi=3\pi/2)$ (in Fig. \ref{figure:sections} c))
and passes ballistically the LB. Apparently, these particles would be entirely injected into 
the chaotic sea of the PSS in the RB once they pass $x_{\text{mid}}$. Due to the negatively directed transport in the RB the particles are transported back to $x_{\text{mid}}$ where they can again be injected into any regular structures 
of the PSS corresponding to the LB, or after several passings of $x_{\text{mid}}$ into regular structures of the RB. Hence, the outgoing particles at $x_{\text{max}}$ for this initially mono energetic beam would occupy all 
accessible regular structures of the PSS of the RB and the corresponding velocity distribution at $x_{\text{max}}$ would contain multiple peaks.\\
As a second example we consider a particle beam (again started at $x_{\text{min}}$) which passes the LB by uniformly occupying the ballistic island at $(v=3.2,\phi=3\pi/2)$. In this case, some particles are injected into 
the ballistic island in the RBs PSS at similar coordinates, while others become diffusive. The particles which remain ballistic traverse the RB and cause a dominant peak in the velocity distribution at $x_{\text{max}}$. For the particles 
which become diffusive the same arguments hold as before. Thus these particles lead to less pronounced peaks in the velocity distribution at velocities corresponding to any kind of regular structure in the RB. 
Accordingly the initial particle beam was converted into a particle beam with a smaller width, because the ballistic island in the LB, i.e. for the initial beam, is larger than the island in the RB in which these particles are injected.
In addition to this, peaks in the velocity distribution emerge due to injection of diffusive particles after multiple crossings of $x_{\text{mid}}$.

%%%%%%%%%%%%%%%%%%%%%%%%%%%%%%%%%%%%%%%%%%%%%%%%%%%%%%%%%%%%%%%%%%%%%%%%%%%%%%%%%%%%%%%%%%%%%%
\subsection{Amplitude variations in superlattices}
\label{section:6.2}

In the following section we demonstrate how the previously discussed interface dynamics of ballistic particles can be exploited to narrow the velocity distribution of particles in an appropriately designed 
superlattice.
To this end we consider a setup build up out of $N_{\text{Bl}}=101$ blocks (whereas each block contains $N=10^4$ barriers) which expands from $x_{\text{min}}=0$ to $x_{\text{max}}=N L  N_{\text{Bl}}$.
Accordingly, the positions of the interfaces, i.e. the positions where the driving laws change, are given by $x_B=N L B$ with $B=1,...,100$.
For the $B=0$ block the driving law is $d_0(t)$ with parameters as in Fig. \ref{figure:sections} c) inducing a positively directed current.
For $B=1,..,100$ the driving laws are: $d_B(t)=A_B[\cos(2.2 t)+\sin(4.4 t)]$ with $A_B=0.3+0.07\cdot B$ (i.e. the 
amplitude is slowly increased from $0.3$ to $1.0$) inducing negatively directed currents. 
The initial conditions for the simulated ensemble are  
$t=0$, $0.4NL<x<0.6NL$ and $-0.1<v<0.1$. 
\begin{figure}[htbp]
\centering
\includegraphics{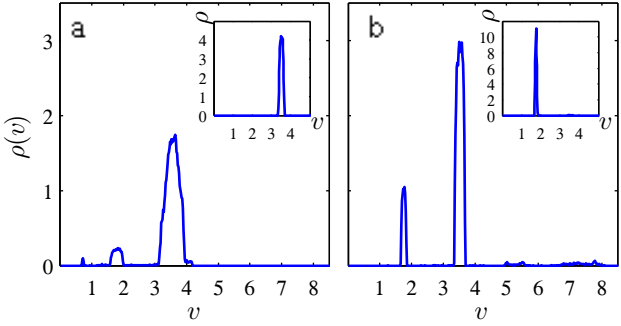}
\caption{\label{peaks}Velocity distributions at positions a) $x_2=2NL$ and b) $x_{\text{max}}=101NL$. Parameters in $d_0(t)$ as in Fig. \ref{density}. For $0<B\leq100$ we set 
$d_B(t)=A_B[\cos(\omega_B t+ \varphi_B)+\sin(2 (\omega_B t+\varphi_B))]$ with $\omega_B=2.2$, $\varphi_B=0$ and $A_B=0.3+0.07(B-1)$. The inset in a) shows $\rho (v)$ at $x=200NL$ with $\omega_B=2.2$, $A_B=A_{101}$ and
  $\varphi_B=\pi(B-1)$ for $100<B\leq200$.
The inset in b) shows $\rho (v)$ at $x=500NL$ with $\omega_B=2.2$, $A_B=A_{101}$ and
  $\varphi_B=0.02\pi(B-1)$ for $100<B\leq500$.} 
\end{figure} 
Hence the particles are located within the chaotic sea of the $B=0$ block with driving law $d_0(t)$ and transported towards the first interface at $x_1=NL$. At this point they can be injected into 
ballistic islands into the phase space in the $B=1$ block. Since the local current in this block is negatively directed, it is hard to surpass for diffusive particles and we obtain a peaked velocity distribution 
\begin{figure}[htbp]
\centering
\includegraphics{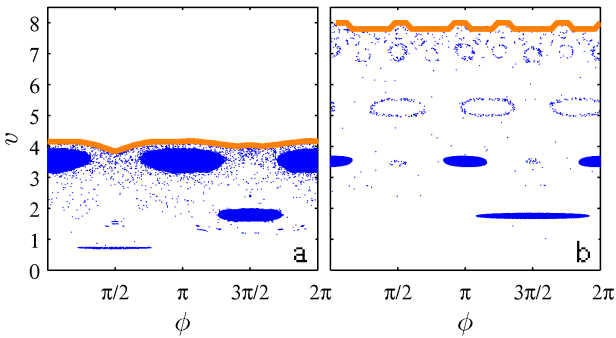}
\caption{\label{nblock_phase}Phase velocity distributions at a) $x=2NL$ and b) $x=101NL$. Orange line indicates the FISC. (Parameters as in Fig. \ref{peaks}).  } 
\end{figure} 
at $x_2=2NL$ (Fig. \ref{peaks} a)) which is   
is dominated by a peak  at $v\approx3.5$ and a less pronounced one at $v\approx 1.8$. The phase velocity distribution at $x_2=2NL$ (Fig. \ref{nblock_phase} a)) reveals that the dominant peak ($v\approx3.5$) can be related 
to an island with winding number $n=2$, while the second peak ($v\approx 1.8$) is associated to a $n=1$ island. \\
In the following blocks the amplitude of the barrier oscillation is subsequently increased and as a result we obtain a velocity distribution with two narrow peaks 
at the same velocities as before for particles at $x_{\text{max}}=101NL$ (Fig. \ref{peaks} b)). Additionally, we observe some particles with velocities $v>4.5$.
The reason for the two dominant peaks is that the amplitude has (as argued in section \ref{section:5}) only little influence on the position 
of ballistic islands in phase space. Thus, most particles remain ballistic at each interface. However the amplitude does have a notable influence on the size of the islands and by choosing the amplitude appropriately, one 
can tune the width of the velocity distribution by adjusting the size of the corresponding ballistic islands. 
In the present setup we exploit that an increasing amplitude leads to a decreasing size of the islands for the used parameters. Hence, the velocity distribution is squeezed when the particles propagate further into the superlattice.
The fast particles with $v>4.5$ correspond to particles in ballistic islands of the underlying phase space which is best seen in the phase velocity distribution at $x=101NL$ (Fig. \ref{nblock_phase} b)). 
In fact, these are the in the previous section discussed peaks in the velocity distribution that emerge due to injection of diffusive particles after multiple crossings of an interface.

%%%%%%%%%%%%%%%%%%%%%%%%%%%%%%%%%%%%%%%%%%%%%%%%%%%%%%%%%%%%%%%%%%%%%%%%%%%%%%%%%%%%%%%%%%%%%%
\subsection{Peaked velocity distributions in superlattices}
\label{section:6.3}

The last step to a mono energetic particle beam is to 
remove one of the peaks in Fig. \ref{peaks} b) without losing too many particles in the other one. 
This can be done by exploiting the symmetries of both islands, which is achieved by adding more blocks to the superlattice with driving laws:  
$d_B(t)=1.0[\cos(2.2 t+\varphi_B)+\sin(4.4 t+2.2\varphi_B)]$ for $101<B<201$ with $\varphi_B=\pi(B-1)$.
Before we show the resulting velocity distributions, let us briefly discuss the idea behind the chosen driving laws:
On the one hand we have seen that
the peak at $v\approx 3.5$ corresponds to a $n=2$ island and consists of two island structures at the same velocity but at different phases (cf. Fig. \ref{nblock_phase}). 
Hence, an additional phase shift $\varphi=\pi$ in the driving law 'maps' both island into each other and most particles remain ballistic. On the other hand the $n=1$ island which is responsible for the peak 
at $v\approx 1.8$ is 'mapped' into the chaotic sea for such a phase shift and particles in it become diffusive. Even though some of these now diffusive particles might be reinjected into a ballistic island of the 
following block, the majority is transported away. Consequently, after performing this procedure multiple times, one obtains a mono energetic particle beam.
The resulting velocity distribution is shown in the inset of Fig. \ref{peaks} a) and reveals that we obtain indeed the desired form of a mono energetic particle beam.\\
At this point we remark that the described technique of removing peaks according to the symmetry of their associated ballistic island works 
for a wide range of different parameter values as well as for ballistic islands with higher winding numbers. 
Unfortunately, it does not apply for the $n=1$ island and thus we can not use it to remove the peak 
at $v\approx 3.5$ while keeping the one at $v\approx 1.8$. 
However, we can exploit that for a large amplitude the $n=1$ island tends to cover a larger range of phases in the PSS (cf. Figs. \ref{phaseplots_omega} g) and h)).
Hence, a small phase shift in the driving law removes relatively fewer particles in the $n=1$ island compared to the ones with higher  $n$. Following this idea, we choose 
$d_B(t)=A_{101}[\cos(2.2 t+\varphi_B)+\sin(4.4 t+2.2\varphi_B)]$ for $101<B<501$ with $\varphi_B=0.02\pi(B-100)$ and the resulting velocity distribution at $x=500NL$ is shown in the inset of Fig. \ref{peaks} b). 
Again, we obtain the desired distribution of a monoenergetic particle beam.

%%%%%%%%%%%%%%%%%%%%%%%%%%%%%%%%%%%%%%%%%%%%%%%%%%%%%%%%%%%%%%%%%%%%%%%%%%%%%%%%%%%%%%%%%%%%%%%%%%%%%%%%%%%%%%%%%%%%%%%%%%%%%%%%%%%%%%%%%%%%%%%%%%%%%%%%%%%%%%%%%%%%%%%%%%%%%%%%%%%%%%%%%%%%%%%%%
\section{Conclusion}   
\label{section:7}

We have explored the classical non-equilibrium dynamics of particles in a one-dimensional driven superlattice which consists of blocks each containing many individual barriers.
While similar systems that are usually studied in this context consist of lattices where all barriers are governed by the same time-dependent force, i.e. driving law,
we allowed for a different driving in each block. 
In doing so we show that the thus obtained variability leads to remarkable new dynamical phenomena. 
To this end we analysed in detail how the blockwise variation of the driving law gives rise to conversion processes from diffusive- to ballistic motion and vice versa at the interfaces, i.e. the positions in the superlattice where the driving 
law changes. 
The combination of directed transport and these conversion processes enabled us to obtain peaked velocity distributions in a simple system containing only two blocks with different 
driving laws providing oppositely directed currents. Additionally, we observed strong correlations between the phases and velocities
for the escaping particles even though the initial particle ensemble is of exclusively diffusive character. 
Even more, we found that the velocity distributions as well as the correlations can be modified in a controlled manner by adjusting parameters such as frequency or amplitude in the driving.
Finally, we present a scheme for superlattices containing a few hundred blocks by witch a diffusive particle ensemble can be converted into a pulsed particle beam, whose mean energy and width in momentum space can be adjusted.
Since this scheme mostly depends on simple 
symmetry arguments it is viable over a wide range of parameters. Thus it should be applicable to 
experimental setups, such as layered semiconductor heterostructures with different AC drivings or even to cold atom experiments in which counter propagating laser beams can create 
a one-dimensional lattice potential. By passing the laser beams through two acousto-optical modulators the desired AC drivings can be obtained. \\
As a future perspective it would be intriguing to explore -both theoretically as well as experimentally- the analogues of the presented effects in the quantum regime.

\section*{Acknowledgments}

C.P. thanks the Excellence Cluster Frontiers in Quantum Photon science, which is supported by the Joachim Herz Stiftung, for financial funding.
We thank F.K. Diakonos for helpful discussions.

\end{document}